\documentclass[aps,prd,showpacs,amsmath,amssymb,twocolumn,superscriptaddress,preprintnumbers,nofootinbib]{revtex4-1}
\pdfoutput=1
\usepackage{graphicx}
\usepackage{amsthm}
\usepackage{amsmath}
\usepackage{graphicx,subfigure}
\usepackage{dcolumn}
\usepackage{bm}
\usepackage{slashed}
\usepackage{calligra}
\usepackage{color}
\DeclareMathAlphabet{\mathcalligra}{T1}{calligra}{m}{n}
\DeclareFontShape{T1}{calligra}{m}{n}{<->s*[2.5]callig15}{}

\newcommand{\be}{\begin{eqnarray}}
\newcommand{\ee}{\end{eqnarray}}
\newcommand{\bea}{\begin{eqnarray}}
\newcommand{\eea}{\end{eqnarray}}

\newcommand{\GeV}{{~\rm GeV}}

\newcommand{\TeV}{{~\rm TeV}}

\newcommand{\ZZ}{{\rm Z}}

\newcommand{\zprime}{Z^\prime}
\newcommand{\mzp}{m_{\zprime}}
\newcommand{\vprime}{v_\Phi}
\newcommand{\gp}{g^\prime}

\begin{document}
\title{Dressing  $L_{\mu}-L_{\tau}$ in Color}
\author{Wolfgang Altmannshofer}
\email{waltmannshofer@perimeterinstitute.ca}
\affiliation{Perimeter Institute for Theoretical Physics 31 Caroline St. N, Waterloo, Ontario, Canada N2L 2Y5.}
\author{Stefania Gori}
\email{sgori@perimeterinstitute.ca}
\affiliation{Perimeter Institute for Theoretical Physics 31 Caroline St. N, Waterloo, Ontario, Canada N2L 2Y5.}
\author{Maxim Pospelov}
\email{mpospelov@perimeterinstitute.ca}
\affiliation{Perimeter Institute for Theoretical Physics 31 Caroline St. N, Waterloo, Ontario, Canada N2L 2Y5.}
\affiliation{Department of Physics \& Astronomy, University of Victoria, Victoria, BC, V8P 5C2, Canada}
\author{Itay Yavin}
\email{iyavin@perimeterinstitute.ca}
\affiliation{Perimeter Institute for Theoretical Physics 31 Caroline St. N, Waterloo, Ontario, Canada N2L 2Y5.}
\affiliation{Department of Physics \& Astronomy, McMaster University 1280 Main St. W. Hamilton, Ontario, Canada, L8S 4L8.}

\begin{abstract}
We consider a new massive vector-boson $Z^\prime$ that couples to leptons through the $L_{\mu}$-$L_{\tau}$ current, and to quarks through an arbitrary set of couplings. We show that such a model can be obtained from a renormalizable field theory involving new heavy fermions in an anomaly-free representation. The model is a candidate explanation for the discrepancy observed recently by the LHCb collaboration in angular distributions of the final state particles in the rare decay $B\rightarrow K^* \mu^+\mu^-$. 
Interestingly, the new vector-boson contribution to the decay $\tau \to \mu \nu_\tau \bar \nu_\mu$ can also remove a small tension in the measurement of the corresponding branching ratio. Constraints from light flavor meson-mixing restrict the coupling to the up- and down-quarks to be very small and thus direct production of the vector-boson at hadron colliders is strongly suppressed.  The most promising ways to test the model is through the measurement of the Z decay to four leptons and through its 
effect on neutrino trident production
of muon pairs. This latter process is a powerful but little-known constraint, which surprisingly rules out explanations of $(g-2)_\mu$ based on $Z^\prime$ gauge bosons coupled to muon number, with mass of at least a few GeV.

\end{abstract}


\maketitle

\section{Introduction}
\label{sec:intro}

Among the indirect probes of weak-scale physics, flavor physics, and $K$, $B$ meson physics in particular, has always played 
a central role. 
The experiments of the last decade have tested the CKM paradigm with impressive accuracy, significantly improving constraints on many classes of physics beyond Standard Model (SM). It is still entirely possible that future improvements in sensitivity to rare decay modes will show a deviation from the SM as the first sign of New Physics (NP) in the flavor sector. 

Recently, a discrepancy in angular observables in the rare decay $B\to K^* \mu^+\mu^-$~\cite{Aaij:2013qta} has motivated several groups~\cite{Descotes-Genon:2013wba,Altmannshofer:2013foa,Beaujean:2013soa,Hurth:2013ssa} to examine NP contributions to the semi-leptonic $b\to s$ currents 
\be
\label{eqn:Hamiltonian}
 \mathcal{H}_\text{eff} = C_9 (\bar s \gamma_\alpha P_L b)(\bar \mu \gamma^\alpha \mu) + C_9^\prime (\bar s \gamma_\alpha P_R b)(\bar \mu \gamma^\alpha \mu) ~.
\ee
It was found that the following choice of parameters gives a particularly good fit to the data~\cite{Altmannshofer:2013foa},
\begin{subequations}
\be
\label{eqn:effective_coeff}
\text{Re}(C_9) &\simeq& -(35~\text{TeV})^{-2} ~,\\ \label{eqn:effective_coeff2}
\text{Re}(C_9^\prime) &\simeq& +(35~\text{TeV})^{-2} ~.
\ee
\end{subequations}
The operators in Eq.~(\ref{eqn:Hamiltonian}) form a subset of a larger family of dimension six semi-leptonic and radiative operators. 
The corresponding axial-vector 
operators $\bar \mu \gamma^\alpha \gamma_5 \mu$ and the magnetic dipole operators $\bar s\sigma^{\mu\nu}F_{\mu\nu}P_{(L,R)}b$ also influence angular observables in the $B$ rare decay. However, the former operator is strongly constrained by the $B_s\to\mu^+\mu^-$ decay, and the latter by $b\to s \gamma$. Therefore, of special interest are the NP models that 
generate the vector coupling to muons, namely $\bar \mu \gamma^\alpha \mu$. In this paper, we adopt the following philosophy: 
it is likely that with more experimental data the existing discrepancy will either be diluted or sharpened; but, in the meanwhile it is reasonable to investigate what classes of NP models the current discrepancy seems to favor. 

The combination of operators in Eq.~(\ref{eqn:Hamiltonian}), together with the absence of axial-vector and magnetic dipole operators, is intriguing as it does not find an immediate home in any well-known extensions of the SM, like the minimal supersymmetric standard model (MSSM) or models with partial compositeness~\cite{Altmannshofer:2013foa}. 
Indeed, if the operators in Eq.~(\ref{eqn:Hamiltonian}) are induced by {\em e.g.} box diagrams of supersymmetric particles, it is more likely that the resulting operators are predominantly coupled to left- or right-handed muons, 
$\bar \mu \gamma^\alpha P_{(L,R)}\mu$. 
Even adjusting the operator chirality structure, LHC bounds on SUSY particle masses strongly constrain the size of NP effects in $B \to K^* \mu^+\mu^-$ to much smaller values compared to Eqs.~(\ref{eqn:effective_coeff},\ref{eqn:effective_coeff2})~\cite{Altmannshofer:2013foa}. 

Nonetheless, a number of NP models have been discussed in the literature~\cite{Gauld:2013qba,Buras:2013qja,Gauld:2013qja,Buras:2013dea,Datta:2013kja,Mahmoudi:2014mja}, mostly based on an additional heavy 
vector boson, or $Z'$. 
To date, several avenues have been investigated with varying degree of model complexity, 
{\em e.g.} a $Z'$ which couples to overall lepton number, and more complete versions with $Z^\prime$ arising from 
the so-called 331 models~\cite{Gauld:2013qba,Buras:2013qja,Gauld:2013qja,Buras:2013dea}. 

In approaching the task of model building, it seems reasonable to first ask what are the models that naturally lead to the ``muon vector portal''  $\bar \mu \gamma^\alpha \mu$? One possibility is a $Z'$ model 
with kinetic mixing between the $Z'$ and hypercharge. 
Unfortunately, such models are subject to strong collider physics constraints. Another possibility is based on gauging the existing approximate global symmetries of the SM. In that respect, one of the most promising candidates is the $U(1)$ gauge group associated with the difference between muon- and tau-lepton number, $L_\mu - L_\tau$, which automatically leads to muonic vector-currents of the required type, $\bar \mu \gamma^\alpha \mu$. 
This gauge group is anomaly-free, and has been the focus of several past phenomenological studies, see {\em e.g.}~\cite{He:1990pn,He:1991qd,Baek:2001kca,Ma:2001md,Salvioni:2009jp,Heeck:2011wj,Feng:2012jn,Harigaya:2013twa}. The absence of anomalies makes this group more attractive than 
gauging overall lepton number, which (if not supplemented by couplings to quarks) is anomalous. 
Gauged $L_\mu - L_\tau$ has all the features of a ``well-hidden'' group: being coupled only to leptons of the second and third generation makes it an extremely difficult 
target for direct collider searches. The $L_\mu - L_\tau$ symmetry has also been studied in the context of neutrino mass model building~\cite{Ma:2001md,Heeck:2011wj}, as it predicts $\theta_{23}\simeq 45^\circ$. 
Finally, as we will also discuss in this paper, it has been shown to accommodate the current discrepancy in $(g-2)_\mu$~\cite{Baek:2001kca,Ma:2001md,Heeck:2011wj,Harigaya:2013twa}. All of the above motivates us to consider an $L_\mu - L_\tau$ portal to $Z'$, and explore its possible connection to rare $B$-decays. 

However, in order to generate the interactions in Eq.~(\ref{eqn:Hamiltonian}), the $Z'$ has to couple to quarks as well, and mediate flavor-changes. 
Given the current data, the pattern associated with the coupling to quarks is unclear. Thus, we choose to remain agnostic about the precise structure associated with the coupling to quarks and follow the ``effective'' approach of ref.~\cite{Fox:2011qd}. 
This allows us to dress the $L_\mu - L_\tau$ vector-boson in color and examine its effects in a fairly general way. At the same time, we provide one concrete realization of UV completion for such couplings via additional heavy vector-like quarks that mix with the SM quarks upon the spontaneous breaking of the $L_\mu - L_\tau$ gauge symmetry.

This paper is organized as follows: in section~\ref{sec:model} we first introduce our ``effective-$Z^\prime$ setup'' and then we discuss a UV-complete version of the model. 
In section~\ref{eqn:flavor}, we derive its consequences for $B$-physics, and determine the parameter range capable of explaining the anomalous measurement from LHCb. Section~\ref{sec:lep} provides an analysis of existing constraints coming from measurements of leptonic processes. 
A particular emphasis is devoted to the constraint coming from the measurement of the neutrino trident production of muon pairs and its role in probing the region of parameter space favored by $(g-2)_\mu$. We close with our conclusions in section~\ref{sec:outlook}.

\section{The Model}
\label{sec:model}

We consider an extension of the SM by a new abelian gauge group, $U(1)^\prime$ (for a review, see for example~\cite{Langacker:2008yv}). Since the associated vector-boson should be massive, we supplement the basic Lagrangian with a scalar field $\Phi$ that ``higgses'' the $U(1)^\prime$, 
\be
\nonumber
\mathcal{L}_{\zprime} &=& -\frac{1}{4}\left(\zprime \right)_{\alpha\beta} \left(\zprime \right)^{\alpha\beta} + \left|D_\alpha \Phi \right|^2 + V(\Phi) 
\\ 
&& + \gp \zprime_\alpha J_{\zprime}^{\alpha} ~.
\ee
Here, $\gp$ is the $U(1)^\prime$ gauge coupling, $J_{\zprime}^{\alpha}$ is the current coupled to the $\zprime$, the field-strength is $\left(\zprime \right)_{\alpha\beta} = \partial_\alpha \zprime_\beta - \partial_\beta \zprime_\alpha$, and $D_\alpha = \partial_\alpha + i \gp \zprime_\alpha$ is the covariant derivative. 
The scalar $\Phi$ has to be charged under the $U(1)^\prime$. We normalize the charge assignments by choosing  $Q_\Phi = +1$. We assume that the potential $V(\Phi)$ is such that the scalar $\Phi$ develops a vacuum expectation value (VEV)
\begin{equation}
\langle \Phi \rangle =  \frac{\vprime}{\sqrt{2}} ~.
\end{equation}
This gives mass to the $\zprime$ gauge boson $\mzp =  \gp \vprime$. The leptonic part of the current is  made purely of $L_\mu - L_\tau$, 
\be
\label{eqn:leptonic_current}
\nonumber
J_{\zprime}^{\alpha~({\rm lep})} &=& Q_{\ell} \Big( \bar{\ell}_2 \gamma^\alpha \ell_2 -\bar{\ell}_3 \gamma^\alpha \ell_3 \\
&& \quad\quad + \bar{\mu}_{_R} \gamma^\alpha \mu_{_R} - \bar{\tau}_{_R} \gamma^\alpha \tau_{_R} \Big)~,
\ee
where $Q_\ell$ is the overall leptonic charge, $\ell_2 = (\nu_\mu , \mu_{_L})$ and $\ell_3 = (\nu_\tau , \tau_{_L})$ are the electroweak doublets associated with left-handed muons and taus, and $\mu_{_R}$ and $\tau_{_R}$ are the right-handed electroweak singlets. $Q_\ell$ 
is a free parameter, but for the rest of the paper we will set $Q_\ell=1$, noting that the dependence of all observables 
on $Q_\ell$ can be easily restored, if needed.

There is rich phenomenology associated with the leptonic current alone as we discuss in the following sections and in 
an upcoming publication~\cite{Altmannshofer:2014pba}. 
However, in this work we are particularly interested in discussing a NP framework able to fit the $B\rightarrow K^* \mu^+\mu^-$ anomaly. We therefore begin by examining the possible couplings of the $Z^\prime$ to quarks and we wish to do so in an as model-independent fashion as possible. The effective-$Z^\prime$ framework of ref.~\cite{Fox:2011qd} (see also~\cite{Carone:2013uh}) is particularly well-suited for this purpose. 
The main observation is that the $Z^\prime$ boson can couple to the quarks through higher dimensional operators suppressed by a NP scale $\Lambda$. In particular, the dimension six operators coupling the quark currents with scalar currents charged under $U(1)'$ contain the $Z'$ field via the covariant derivative:
\be \label{eq:penguin} 
\mathcal{L}_{\rm dim6} &=& (\Phi^* i \overleftrightarrow{D_\alpha} \Phi) \Bigg[ \frac{\lambda_{ij}^{(q)}}{\Lambda^2} (\bar q_{_L}^i \gamma^\alpha q_{_L}^j) \nonumber \\
&& + \frac{\lambda_{ij}^{(d)}}{\Lambda^2} (\bar d_{_R}^i \gamma^\alpha d_{_R}^j) + \frac{\lambda_{ij}^{(u)}}{\Lambda^2} (\bar u_{_R}^i \gamma^\alpha u_{_R}^j) \Bigg] ~,
\ee
where $\Phi^* i \overleftrightarrow{D_\alpha} \Phi = -i (D_\alpha \Phi)^* \Phi + i \Phi^*(D_\alpha \Phi)$, $q_{_L}=(u_{_L},d_{_L})$ and $d_{_R},\,u_{_R}$ are the quark $SU(2)_L$ doublet and singlets, respectively. Generically, the couplings $\lambda_{ij}^{(q),(u),(d)}$ are complex $3\times 3$ matrices with $\mathcal{O}(1)$ entries.
After the $U(1)^\prime$ symmetry is spontaneously broken by the VEV of $\Phi$, the operators in Eq.~(\ref{eq:penguin}) result in effective couplings of the SM quarks to the $\zprime$. The hadronic part of the $U(1)^\prime$ current is then given by,
\be
\label{eqn:hadronic_current}
\nonumber
J_{\zprime}^{\alpha~({\rm had})}  &=&  \mathbb{R}^{(d)}_{ij} \bar{d}_i \gamma^\alpha P_{_R} d_j  + \mathbb{L}^{(d)}_{ij} \bar{d}_i \gamma^\alpha P_{_L} d_j
\\
&+& \mathbb{R}^{(u)}_{ij} \bar{u}_i \gamma^\alpha P_{_R} u_j  + \mathbb{L}^{(u)}_{ij} \bar{u}_i \gamma^\alpha P_{_L} u_j~,
\ee
where $d_i$ is a down-type quark mass eigenstate of flavor $i$, $u_j$ is an up-type quark of flavor $j$, and $P_{_L}$ and $P_{_R}$ are the left and right-handed Dirac projection operators. We find
\begin{subequations}
\be
 \mathbb{R}^{(d)}_{ij} &=& \lambda_{ij}^{(d)} \frac{v_\Phi^2}{\Lambda^2}~,~~
 \mathbb{R}^{(u)}_{ij} = \lambda_{ij}^{(u)} \frac{v_\Phi^2}{\Lambda^2} ~,
 \\
 \mathbb{L}^{(d)}_{ij} &=& \lambda_{ij}^{(q)} \frac{v_\Phi^2}{\Lambda^2} ~,~~
\mathbb{L}^{(u)}_{ij} = (V \mathbb{L}^{(d)} V^\dagger)_{ij} ~,
\ee
\end{subequations}
where $V$ is the Cabibbo-Kobayashi-Maskawa (CKM) quark mixing matrix.

\begin{figure*}[t]
\centering
\includegraphics[width=0.24\textwidth]{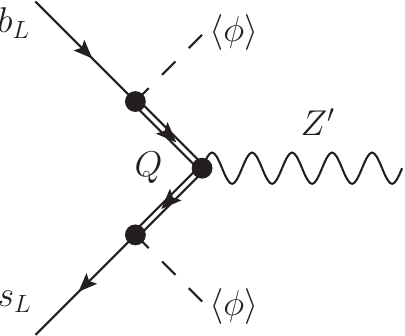} ~~~~~~~
\includegraphics[width=0.24\textwidth]{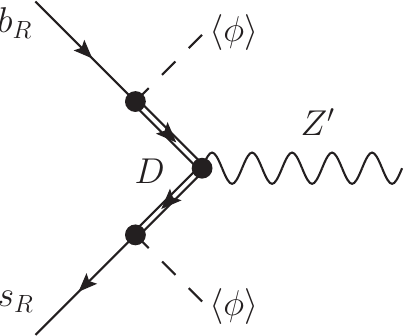} ~~~~~~~
\includegraphics[width=0.24\textwidth]{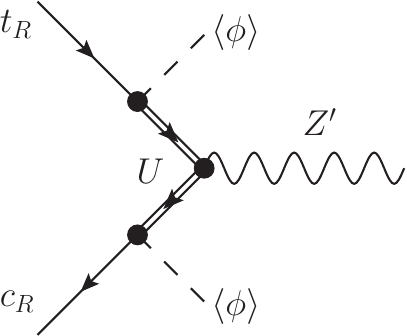}
\caption{Example diagrams in the high energy theory that lead to flavor-changing effective couplings of the $Z^\prime$ to SM quarks.}
\label{fig:effective_couplings}
\end{figure*}

In the rest of this section, we follow ref.~\cite{Fox:2011qd} and show that, starting with an anomaly-free renormalizable field theory at high-energy, we can obtain a fairly general structure for the $U(1)^\prime$ coupling to quarks at low energies. We introduce one set of vector-like (heavy) fermions $Q$, $D$, $U$, that are copies of the SM quarks, but are charged under the new $U(1)^\prime$: $Q_Q = Q_\Phi =  - Q_D = - Q_U = 1$. The vector-like heavy quarks decompose as,
\begin{subequations}
\be \label{eq:9a}
 Q_{_L}  &=& (\mathbf{3}, \mathbf{2})_{+1/6, +1} ~,~~ \tilde Q_{_R}  = (\mathbf{3}, \mathbf{2})_{+1/6, +1} ~, \\  \label{eq:9b}
 \tilde D_{_L} &=& (\mathbf{3}, \mathbf{1})_{-1/3, -1} ~,~~ D_{_R} = (\mathbf{3}, \mathbf{1})_{-1/3, -1} ~, \\
 \tilde U_{_L} &=& (\mathbf{3}, \mathbf{1})_{+2/3, -1} ~,~~ U_{_R} = (\mathbf{3}, \mathbf{1})_{+2/3, -1} ~,
\ee
\end{subequations}
where we denote the weak doublets as $Q_{_L}= (U_{_L}, D_{_L})$ and $\tilde Q_{_R} = (\tilde U_{_R}, \tilde D_{_R})$. In addition to the vector-like mass terms,
\be
\mathcal{L}_m = m_Q \bar Q_{_L} \tilde Q_{_R} + m_D \bar{\tilde D}_{_L} D_{_R} +  m_U \bar{\tilde U}_{_L} U_{_R} ~+~ \text{h.c.} ~,
\ee
the above quantum number assignments allow the following Yukawa couplings that mix the vector-like quarks with the SM quarks, after $\Phi$ gets a VEV:
\be
\mathcal{L}_{\rm mix} &=& \Phi \bar{\tilde D}_{_R} (Y_{Qb} b_{_L} + Y_{Qs} s_{_L} + Y_{Qd} d_{_L}) 
\\
&+& \Phi \bar{\tilde U}_{_R} (Y_{Qt} t_{_L} + Y_{Qc} c_{_L} + Y_{Qu} u_{_L}) \nonumber
\\
&+& \Phi^\dagger \bar{\tilde U}_{_L} (Y_{Ut} t_{_R} + Y_{Uc} c_{_R} + Y_{Uu} u_{_R}) \nonumber
\\
&+& \Phi^\dagger \bar{\tilde D}_{_L} (Y_{Db} b_{_R} + Y_{Ds} s_{_R} + Y_{Dd} d_{_R})~+~ \text{h.c.} ~, \nonumber
\ee
where $Y_{Qb}$, for example, denotes the Yukawa coupling associated with the mixing between the field $Q$ and the SM left-handed field $b_{_L}$. Electroweak invariance forces the relation $(Y_{Qu},Y_{Qc},Y_{Qt}) = V^* (Y_{Qd},Y_{Qs},Y_{Qb})$. 
Throughout we work in the CKM basis, where all SM quark Yukawas are diagonal. Note that, due to our choice of $U(1)^\prime$ charges for the vector-like quarks, no couplings of the vector-like quarks with the SM Higgs are possible.

Below the energy scale where the scalar $\Phi$ acquires a VEV, the above couplings lead to the following quark mass matrices,
\begin{subequations}
\be
\mathcal{M}_u = \begin{pmatrix} 
m_Q & 0 & Y_{Qt} \frac{v_\Phi}{\sqrt{2}} & Y_{Qc} \frac{v_\Phi}{\sqrt{2}} & Y_{Qu} \frac{v_\Phi}{\sqrt{2}} \\ 
0 & m_U & 0 & 0 & 0 \\
0 & Y_{Ut} \frac{v_\Phi}{\sqrt{2}} & m_t & 0 & 0 \\
0 & Y_{Uc} \frac{v_\Phi}{\sqrt{2}} & 0 & m_c & 0 \\
0 & Y_{Uu} \frac{v_\Phi}{\sqrt{2}} & 0 & 0 & m_u
 \end{pmatrix} ~,
\ee
\be
\mathcal{M}_d = \begin{pmatrix} 
m_Q & 0 & Y_{Qb} \frac{v_\Phi}{\sqrt{2}} & Y_{Qs} \frac{v_\Phi}{\sqrt{2}} & Y_{Qd} \frac{v_\Phi}{\sqrt{2}} \\ 
0 & m_D & 0 & 0 & 0 \\
0 & Y_{Db} \frac{v_\Phi}{\sqrt{2}} & m_b & 0 & 0 \\
0 & Y_{Ds} \frac{v_\Phi}{\sqrt{2}} & 0 & m_s & 0 \\
0 & Y_{Dd} \frac{v_\Phi}{\sqrt{2}} & 0 & 0 & m_d
 \end{pmatrix} ~.
\ee
\end{subequations}
Diagonalizing these mass matrices results in small relative corrections to the mass eigenvalues of the vector-like quarks and SM quarks of order $v_\Phi^2/m_{Q,U,D}^2$, as well as to small corrections to the CKM angles. While the couplings of the SM Higgs boson to SM fermions are {\it not} affected in this framework, the mixing with the vector-like quarks does result in modification of the couplings of the $\ZZ$ boson to the quarks of the SM. 
However, due to the $SU(2)_L$ symmetry, these corrections are necessarily suppressed by SM quark masses and are negligibly small for vector-like quark masses above the TeV scale. Furthermore, due to the unbroken $U(1)_\text{em}$, the photon does not acquire flavor changing couplings.

The main effect of the mixing of SM and vector-like quarks is to generate effective couplings of the SM quarks to the $\zprime$ as defined in Eq.~(\ref{eq:penguin}) and illustrated by the Feynman diagrams in Fig.~\ref{fig:effective_couplings}. Matching it to 
Eq.~(\ref{eq:penguin}), we find
\begin{subequations}
\be
\lambda_{ij}^{(d)} \frac{1}{\Lambda^2} &=& -\frac{(Y_{Di} Y_{Dj}^*)}{2 m_D^2} ~, \\
\lambda_{ij}^{(u)} \frac{1}{\Lambda^2} &=& -\frac{(Y_{Ui} Y_{Uj}^*)}{2 m_U^2}~, \\
\lambda_{ij}^{(q)} \frac{1}{\Lambda^2} &=& \frac{(Y_{Qi} Y_{Qj}^*)}{2 m_Q^2}~.
\ee
\end{subequations}

Before closing this section, we note that, the leptonic current of Eq.~(\ref{eqn:leptonic_current}), and the hadronic current of Eq.~(\ref{eqn:hadronic_current}) completely determine the leading order low-energy effects of the $\zprime$ vector-boson. In particular, no coupling to the electron is present at this level. 
However, quantum corrections due to the heavy exotic quarks can lead to kinetic mixing between the $\zprime$ and the gauge-boson of hypercharge as in,
\be
\label{eqn:kinetic_mixing}
\mathcal{L}\supset -\frac{\epsilon}{2} \zprime_{\alpha\beta} B^{\alpha\beta}~.
\ee
As is well-known, this mixing leads to a coupling of the $\zprime$ with all fields that carry hypercharge, hence to the entire SM matter content. In the renormalizable model presented above, the induced mixing is UV finite, due to the chosen $U(1)^\prime$ charges. Close to the vector-like quark threshold, the mixing is given by 
\be
\epsilon = \frac{g^\prime g_1}{16 \pi^2} \frac{2}{9} \log\left( \frac{m_Q^2 m_D^2}{m_U^4} \right) ~.
\ee
Being loop suppressed, unless the spectrum of the vector-like quarks is very hierarchical, this kinetic mixing is typically below the $10^{-3}$ level. Finally, we have to mention that additional effects on the $Z^\prime$ phenomenology could arise from mixing of 
the Higgs boson with the scalar $\Phi$ breaking the $U(1)^\prime$ symmetry, for example through the Higgs portal operator $|H|^2|\Phi|^2$. The effects, however, are more model dependent and we do not study them in this work.

\section{The  {\boldmath $B \to K^* \mu^+\mu^-$} anomaly and additional flavor constraints}
\label{eqn:flavor}

Before discussing the various constraints on the hadronic current of Eq.~(\ref{eqn:hadronic_current}), we match the Wilson coefficients relevant for the $B \to K^* \mu^+\mu^-$ anomaly, Eqs.~(\ref{eqn:effective_coeff},\ref{eqn:effective_coeff2}) with the corresponding terms in the $\zprime$ currents. 
Working in the approximation that the $\zprime$ is heavy compared to the $B$ meson\footnote{If the $\zprime$ is lighter than the $B$ meson, it would show up as a resonance in the di-muon invariant mass spectrum of the $B \to K^* \mu^+\mu^-$ decay rate. We reserve the analysis to another publication~\cite{Altmannshofer:2014pba}.}, so as to neglect the momentum exchange in the semi-leptonic decay of the B, we have
\begin{subequations}
\be
\label{eq:anomaly_C9}
C_9 &=& \lambda_{bs}^{(q)} \frac{1}{\Lambda^2} = \frac{Y_{Qb} Y_{Qs}^*}{2m_Q^2}
~, \\
\label{eq:anomaly_C9p}
C_9^\prime &=& \lambda_{bs}^{(d)} \frac{1}{\Lambda^2} =-\frac{Y_{Db} Y_{Ds}^*}{2m_D^2} 
~,
\ee
\end{subequations}
with the relative minus sign arising from the opposite $U(1)^\prime$ charges of $\tilde Q_{_R}$ and $\tilde D_{_L}$ (see Eqs.~(\ref{eq:9a},\ref{eq:9b})). 
We note that in this approximation the Wilson coefficients $C_9$ and $C_9^\prime$ are completely independent of the $Z^\prime$ mass and the $U(1)^\prime$ gauge coupling. Therefore, these relations determine the mass scale for the exotic quarks, 
\be
m_{Q,D} \simeq 25 \TeV \times \left( {\rm Re} (Y_{(Q,D)b} Y_{(Q,D)s}^*) \right)^{1/2} ~,
\ee
in order to address the anomaly in the $B \to K^* \mu^+\mu^-$ decay (see Eqs.(\ref{eqn:effective_coeff},\ref{eqn:effective_coeff2})).
This scale is sufficiently high that current collider constraints on new colored particles ($\gtrsim 1\TeV$) do not result in useful bounds. However, other flavor processes are easily sensitive to such high scales. While they do not rule out the combinations leading to the operators corresponding to $C_9$ and $C_9^\prime$, they do place constraints on the general mixing coefficients as we now discuss. 

\begin{figure*}[t]
\centering
\includegraphics[width=0.46\textwidth]{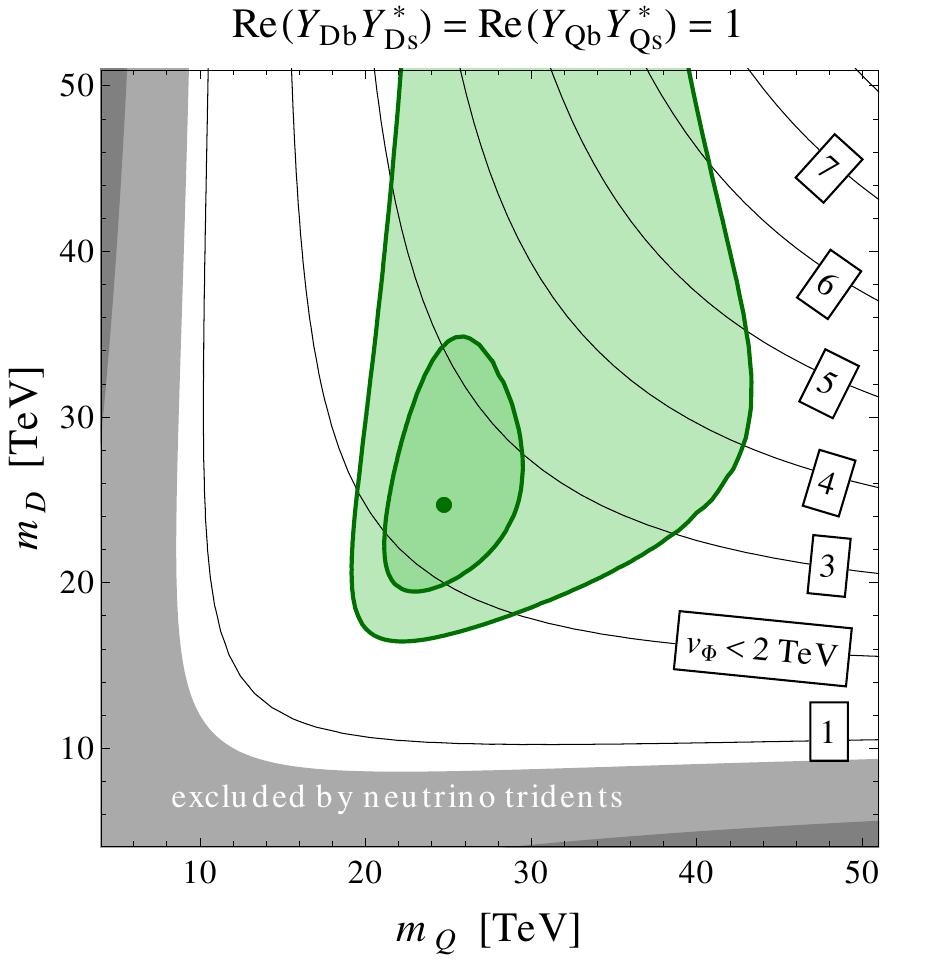} ~~~~~
\includegraphics[width=0.46\textwidth]{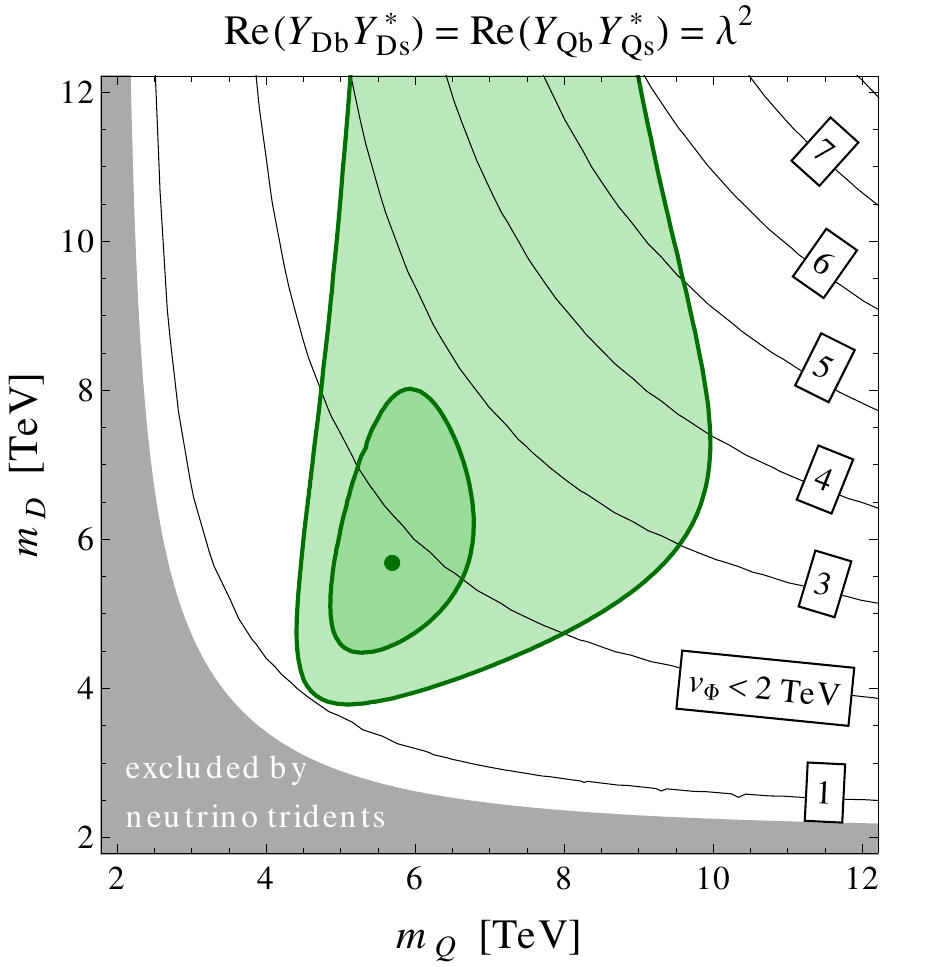}
\caption{Constraints from $B_s$ mixing on the $U(1)^\prime$ breaking VEV, $v_\Phi$, in the plane of the vector-like quark masses $m_Q$ and $m_D$. In the left plot all relevant mixing Yukawas are set to 1. In the right plot, we assume a non trivial flavor structure that leads to $Y_{Qs} \simeq Y_{Ds} \simeq \lambda^2$, where $\lambda \simeq 0.23$ is the Cabibbo angle. The green region is preferred by an explanation of the $B \to K^* \mu^+\mu^-$ anomaly. The light gray regions are excluded by experimental results on neutrino trident production (see Eq.~(\ref{eq:trident_bound}) below). The dark gray region in the left plot cannot be made compatible with $B_s$ mixing bounds.}
\label{fig:mixing}
\end{figure*}

{\bf Meson mixing:} Tree level exchange of the $\zprime$ contributes to neutral meson mixing.
In particular, the couplings required to explain the $B \to K^* \mu^+\mu^-$ anomaly will lead to contributions to $B_s$ mixing. 
Additional contributions to $B_s$ mixing arise from the flavor-changing effects associated with the scalar $\Phi$. Both real and imaginary parts of 
$\Phi$ (the latter is equivalent to the longitudinal part of the $Z'$) mediate SM$-$vector-like quark transitions, and the 
box diagram with $\Phi$ exchange therefore leads to an additional contribution to $\Delta B= 2$ transitions.

The modifications to the mixing amplitude $M_{12}$ read
\begin{eqnarray} \label{eq:M12_box}
\frac{M_{12}}{M_{12}^\text{SM}} &=& 1 + \Big[ C_{LL} + C_{RR} + 9.7 C_{LR} \Big] \nonumber \\
&& \times \left( \frac{g_2^4}{16\pi^2} \frac{1}{m_W^2} (V_{ts}^*V_{tb})^2 S_0 \right)^{-1} ~,
\end{eqnarray} 
where we used the hadronic matrix elements collected in~\cite{Buras:2012jb}, and the SM loop function is $S_0 \simeq 2.3$. 
The Wilson coefficients $C_{LL},C_{RR},C_{LR}$ are given by 
\begin{subequations}
\begin{eqnarray}
C_{LL} &=& (Y_{Qb}Y_{Qs}^*)^2 \left( \frac{v_\Phi^2}{m_Q^4} + \frac{1}{16\pi^2} \frac{1}{m_Q^2} \right) ~,\\[8pt]
C_{RR} &=& (Y_{Db}Y_{Ds}^*)^2 \left( \frac{v_\Phi^2}{m_D^4} + \frac{1}{16\pi^2} \frac{1}{m_D^2} \right) ~,\\[8pt]
C_{LR} &=& (Y_{Qb}Y_{Qs}^*)(Y_{Db}Y_{Ds}^*) \nonumber \\
&& \times \left( \frac{v_\Phi^2}{m_Q^2m_D^2} - \frac{1}{16\pi^2} \frac{\log(m_Q^2/m_D^2)}{m_Q^2 - m_D^2} \right) ~, \label{eq:19c}
\end{eqnarray} 
\end{subequations}
where the $\mathcal{O}(v_\Phi^2)$ terms originate from tree level $Z'$  contributions, and the $1/(16\pi^2)$ suppressed contributions 
originate from the scalar box diagrams. 
Note that the $Z^\prime$ contribution to the mixing amplitude does not depend on the $Z^\prime$ mass and the $U(1)^\prime$ gauge couplings separately, but only through the combination $v_\Phi = \mzp / g^\prime$. The good agreement of the SM prediction for $B_s$ mixing with the experimental data sets an {\it upper} bound on the $U(1)^\prime$ symmetry breaking VEV, $v_\Phi$. 

In the plots of Fig.~\ref{fig:mixing} we show the limit on $v_\Phi$ as a function of the masses of the vector-like quarks, $m_D$ and $m_Q$. In the plot on the left-hand side we fix the Yukawa couplings Re$(Y_{Db} Y_{Ds}^*)={\rm{Re}}(Y_{Qb} Y_{Qs}^*) =1$.
The region favored by an explanation of the $B \to K^* \mu^+\mu^-$ anomaly is shaded in green at the 1$\sigma$ and 2$\sigma$ level. The dark gray region is excluded by $B_s$ mixing constraints\footnote{Uncertainties in the SM prediction, coming mainly from the limited precision of the hadronic matrix elements and CKM factors, allow for modest NP effect in $B_s$ mixing. In the plot we allow for NP effects of up to $\sim 15\%$.}. The light gray region is excluded by neutrino tridents (see Eq.~(\ref{eq:trident_bound}) below). If the Yukawas are of order unity, the region best fitting the $B \to K^* \mu^+\mu^-$ anomaly corresponds to vector-like masses of the order of 25 TeV and the scalar box contributions to $B_s$ mixing are sizable.

Next, we assume a flavor hierarchy in the Yukawa couplings associated with mixing,
\be
\label{eqn:mixing_structure}
\frac{Y_{Qs}}{Y_{Qb}}&\sim&\frac{Y_{Ds}}{Y_{Db}}\sim \lambda^2 ~,
\ee
with the Cabibbo angle $\lambda \simeq 0.23$.
In this case, as shown in the right-hand plot of Fig.~\ref{fig:mixing}, relatively light vector-like quarks of the order of 5 TeV are required for the $Z^\prime$ to explain the $B \to K^* \mu^+\mu^-$ anomaly. That implies that the $Z^\prime$ contributions to $B_s$ mixing dominate and scalar box contributions can be neglected. To avoid the experimental constraint from $B_s$ mixing then requires
\begin{equation}\label{eq:18TeVbound}
v_\Phi \lesssim 1.8 \TeV \quad \Rightarrow \quad \mzp \lesssim \gp\cdot 1.8 \TeV ~.
\end{equation}
A mild prior on the $U(1)^\prime$ gauge coupling $g^\prime$ comes from the requirement of 
the gauge coupling to remain perturbative all the way to the Planck energy scale and we find $g^\prime \lesssim 0.35$. 
Thus, we expect such a $\zprime$ vector-boson that explains the $B \to K^* \mu^+\mu^-$ anomaly to be below a TeV. 

The constraints from other flavor-changing processes are more model dependent. In particular, strong constraints on the combinations of the heavy exotic quark parameters involving first and second generations can be derived from Kaon mixing. In the presence of an $\mathcal{O}(1)$ phase in flavor-changing couplings to both left-handed and right-handed quarks, CP violation in Kaon mixing leads to a particular strong bound. 
Neglecting $\Phi$-box contributions, and using bounds given in~\cite{Isidori:2010kg,Butler:2013kdw}, we find
\be
{\rm Im}(\lambda_{sd}^{(q)}\lambda_{sd}^{(d)}) \frac{2 v_\Phi^2}{\Lambda^4} &=& {\rm Im}(Y_{Qs}Y_{Qd}^*Y_{Ds}Y_{Dd}^*) \frac{v_\Phi^2}{2m_Q^2m_D^2} \nonumber \\
&& \lesssim (3.2 \times 10^{5}~{\rm TeV})^{-2}~.
\ee
In order to obtain bounds on $\vprime$, we need to specify the Yukawa couplings that are responsible for mixing with the first generation. A consistent extension of the flavor structure in~(\ref{eqn:mixing_structure}) is
\be
\label{eqn:mixing_structure_db}
\frac{Y_{Qd}}{Y_{Qb}}&\sim&\frac{Y_{Dd}}{Y_{Db}}\sim \lambda^3 ~.
\ee
This structure strongly suppresses the $\zprime$ contributions to Kaon mixing, and $\vprime \lesssim 350\GeV$ would satisfy all Kaon mixing constraints together with the $B\to K^* \mu^+\mu^-$ anomaly. However, as we will see in the following, such a small VEV is ruled out, primarily by neutrino trident production (see Eq.~(\ref{eq:trident_bound}) below). 

The least constrained possibility is when the coupling to the first generation quarks is entirely suppressed. While it is possible to simultaneously set to zero the couplings to the first generation of right-handed up and down quarks, $SU(2)_L$ invariance implies $Y_{Qu} = Y_{Qd} + \lambda Y_{Qs}$ at leading order in the Cabibbo angle. That means that NP effects in Kaon mixing and in charm mixing cannot be switched off simultaneously. However, we checked explicitly that switching off NP in Kaon mixing, $B_s$-mixing still gives the strongest constraint and $\vprime \lesssim 1.8\TeV$ as derived above, in Eq.~(\ref{eq:18TeVbound}).

{\bf Other searches:} We close this section with a brief discussion of other possible searches in hadronic processes motivated by the $\zprime$ we consider in this work. First, we note that the possibilities discussed above require very weak effective couplings of the $\zprime$ with the first generation quarks. 
The resulting cross-section for direct production of the $\zprime$ in hadronic colliders is therefore too small to be observed, even in the most sensitive leptonic resonance searches.

The mixing with the second and third generation as in Eq.~(\ref{eqn:mixing_structure}) can possibly be searched for in heavy flavor processes. However, the most obvious possibility, namely the $B_s \to \mu^+\mu^-$ decay, remains SM-like in our framework since muonic axial-vector currents are absent. Furthermore, also $b\to s \gamma$ does not receive relevant contributions, as NP effects are only induced at the loop level.
Other rare $B$ meson decays that are based on the $b \to s$ transition, such as the inclusive $B \to X_s \ell^+\ell^-$ processes, 
can be affected by the $Z^\prime$. In particular, while the electron mode, $B \to X_s e^+e^-$ remains SM-like, our framework predicts a $\sim 20\%$ suppression (enhancement) of the muonic (tauonic) mode $B \to X_s \mu^+\mu^-$ ($B \to X_s \tau^+\tau^-$). Such modifications are interesting goals for Belle II~\cite{Aushev:2010bq}.
Modifications of the neutrino modes $B \to X_s \nu\bar\nu^-$, $B \to K \nu\bar\nu^-$, and $B \to K^* \nu\bar\nu^-$, on the other hand, will be challenging to observe. As the neutrino flavor cannot be observed in the experiment, the NP effects cancel at leading order and the decay rates are only marginally increased by about $\sim 2\%$. We note in passing, that, although we cannot predict the magnitude of the 
modification for the $K^+\to \pi^+ \nu\bar\nu$, the $L_\mu-L_\tau$-based $Z'$ always {\em increases} the decay rate
regardless of the sign of the effective $s-d-Z'$ vertex, and $\sim O(10\%)$ modification could become observable at the next installment of this 
search~\cite{Kozhuharov:2013oca}. 

Another possibility, although likely too small, is rare decays of the top into up or charm quarks and a gauge boson. If kinematically allowed, the $t \to \zprime c$ branching ratio is
\be
\label{eqn:tZpc_BR}
\text{BR}(t &\to&  \zprime c) \simeq \frac{2(1-x^\prime)^2(1+2x^\prime)}{(1-x)^2(1+2x)} \\ \nonumber
&\times& \left(|Y_{Qt} Y_{Qc}^*|^2 \frac{v^2 v_\Phi^2}{4m_Q^4} + |Y_{Ut} Y_{Uc}^*|^2 \frac{v^2 v_\Phi^2}{4m_U^4}\right) ~,
\ee
with
\be
x = \frac{m_W^2}{m_t^2} ~,~~ x^\prime = \frac{\mzp^2}{m_t^2} ~. 
\ee
The first term in the parenthesis is required to explain the $B \to K^* \mu^+\mu^-$ anomaly. Again, due to $SU(2)_L$ invariance, at leading order, we have $Y_{Qt}\sim Y_{Qb}$ and $Y_{Qc}\sim Y_{Qs}$. Using only this term with $ {\rm{Re}}(Y_{Qt} Y_{Qc}^*)/(2 m_Q^2) \simeq 1/(35\,{\rm{TeV}})^2$, as required by Eqs. (\ref{eqn:effective_coeff}), (\ref{eq:anomaly_C9}), and the upper bound on $v_\Phi$ from $B_s$ meson mixing ($v_\Phi \lesssim 1.8 \TeV$), we find branching ratios of at most few$\times 10^{-7}$. We note that the second term in the parenthesis is  connected to the right-handed up quark sector and therefore unrelated to the $B$ physics phenomenology. In principle this second term could be larger, leading to branching ratios as large as $1\%$, for vector-like masses of $\mathcal{O}(1\TeV)$ and mixing Yukawas of $\mathcal{O}(1)$. ATLAS and CMS collaborations both search for the rare decay $t\to c Z$~\cite{Aad:2012ij,Chatrchyan:2013nwa}. In particular, CMS~\cite{Chatrchyan:2013nwa} sets the most 
stringent bound BR$(t \to c Z) < 5\times 10^{-4}$, analyzing the full 7+8 TeV data set. The expected reach of the 14 TeV LHC with 300 fb$^{-1}$ data is at the level of $10^{-5}$~\cite{ATLAS:2013hta,CMS:2013xfa}. To set the present bound, the CMS collaboration investigates the process $pp\to t\bar t\to (Wb) (cZ)$ with both $\ZZ$ and $W$ decaying leptonically and with two leptons forming an invariant mass compatible with a $\ZZ$ boson. Our $Z^\prime$ candidate has a 1/3 branching ratio to a pair of muons. Therefore, a search for $t\to \zprime c$ could be done in a similar fashion by relaxing the cut on the lepton invariant mass. 

The branching ratios of other rare top decays, in particular $t \to c Z$, $t \to c h$ and the radiative decays $t \to c \gamma$ and $t \to c g$ are smaller than BR$(t \to c Z^\prime)$ by roughly a loop factor.

\section{Constraints from high precision leptonic processes} \label{sec:lep}

The coupling of the $\zprime$ to leptons is restricted to take the form Eq.~(\ref{eqn:leptonic_current}) by the gauge symmetry $L_{\mu}-L_{\tau}$. Therefore, the contributions to the leptonic phenomenology depend only on the gauge coupling and mass of the $\zprime$ (aside from small, model-dependent corrections associated with the kinetic mixing, Eq.~(\ref{eqn:kinetic_mixing})). In this section, we discuss the effects of the $\zprime$ on the muonic $g-2$ value, the tau leptonic decay width, the leptonic widths of the SM $\ZZ$, and searches for 4 lepton events at colliders. Finally, we discuss a new and important constraint coming from the observation of neutrino trident production.

\begin{figure}[t]
\centering
\includegraphics[width=0.48\textwidth]{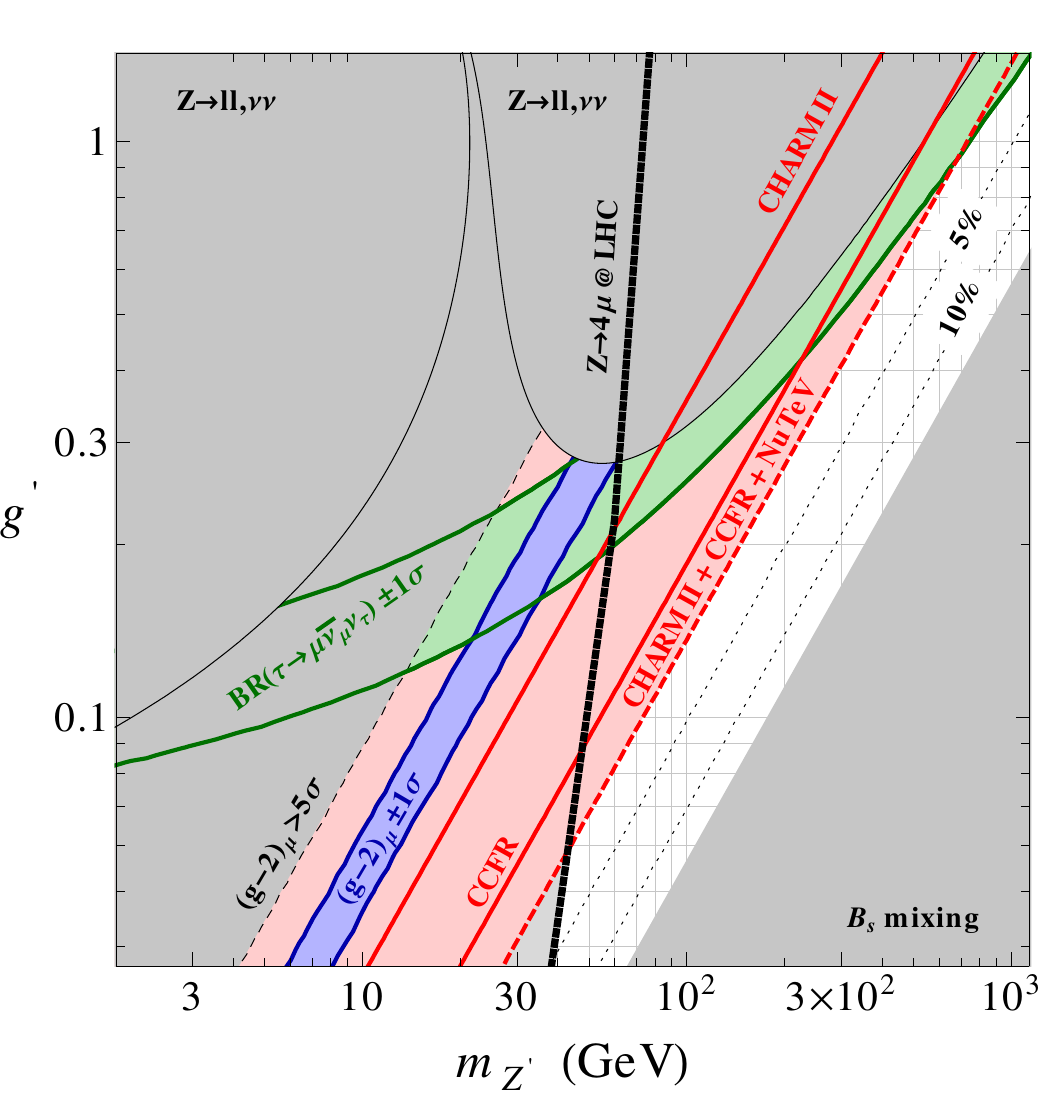}
\caption{Constraints on the model parameter space from the different leptonic processes discussed in Section~\ref{sec:lep}. The region in white is the allowed region. The anomaly in $B\rightarrow K^* \mu^+\mu^-$ can be accommodated everywhere to the left of the bottom-right triangle, see Eq.~(\ref{eq:18TeVbound}). Note that the constraint from the neutrino trident production of muon pairs (red region) completely excludes the region favored by $(g-2)_\mu$. The dotted lines in the allowed region denote $(5-10)\%$ NP effects in $B_s$ mixing.}
\label{fig:constraints1}
\end{figure}

\bigskip

{$\bullet$ {\boldmath $(g-2)_\mu$.} At one-loop level the $\zprime$ contribution to the muon $g-2$~\cite{Pospelov:2008zw} reads
\begin{equation}
\label{eqn:muonic_g-2}
 \Delta a_\mu = \frac{1}{12\pi^2} \frac{m_\mu^2}{v_\Phi^2} ~,
\end{equation}
where we assumed that $\mzp \gg m_\mu$. In this limit, the contribution depends only on the $U(1)^\prime$ symmetry breaking VEV. Given the well-known discrepancy between theory and experiment, an additional contribution of $\Delta a_\mu = (2.9 \pm 0.9) \times 10^{-9}$ to the theoretical value would be required
~\cite{Jegerlehner:2009ry}. In our model, this determines the VEV to be $v_\Phi \simeq 180\GeV$. The corresponding $1\sigma$ range is shown in Fig.~\ref{fig:constraints1} as the blue diagonal band. Alternatively, this measurement sets a $\sim 5\sigma$ lower bound on the VEV of $v_\Phi \gtrsim 110\GeV$ such that $\Delta a_\mu\lesssim 7.4\times 10^{-9}$ (see the diagonal gray region in Fig.~\ref{fig:constraints1}).


\begin{figure}[t]
\centering
\includegraphics[width=0.4\textwidth]{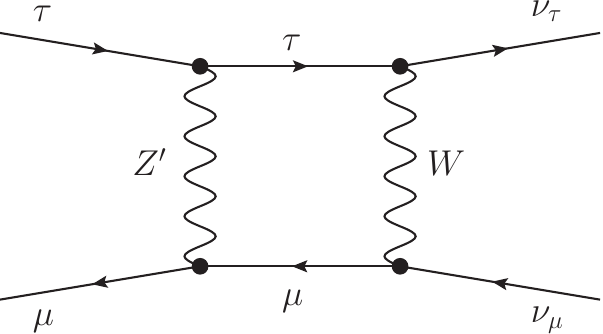}
\caption{Example one-loop box diagram that gives a correction to the $\tau \to \mu \nu_\tau \bar \nu_\mu$ decay. In total there are four box diagrams with the $Z^\prime$ connected to the lepton legs.}
\label{fig:tau_diagram}
\end{figure}
\bigskip
{$\bullet$ {\boldmath $\tau$} {\bf{decays.}}} The $\zprime$ also leads to corrections to tau decay processes. In particular, one-loop box diagrams, such as the one shown in Fig.~\ref{fig:tau_diagram}, give the leading modifications to the $\tau \to \mu \nu_\tau \bar \nu_\mu$ rate, while the $\tau \to e \nu_\tau \bar \nu_e$ decay remains SM-like to an excellent approximation. Contributions to $\tau \to e \nu_\tau \bar \nu_e$ (and $\tau \to \mu \nu_\tau \bar \nu_\mu$) from vertex corrections are suppressed by a factor $m_\tau^2/\mzp^2$ due to $SU(2)_L$ invariance and can be safely neglected in the regions of parameter space we are interested in. Tiny additional corrections can arise in the presence of kinetic $Z-Z^\prime$ mixing.  
Evaluating the box diagrams, we find the following correction
\be
 \frac{\text{BR}(\tau \to \mu \nu_\tau \bar \nu_\mu)}{\text{BR}(\tau \to \mu \nu_\tau \bar \nu_\mu)_\text{SM}} \simeq  1 + \Delta ~,
\ee
where,
\be
\Delta = \frac{3 (\gp)^2}{4\pi^2} \frac{ \log(m_W^2/\mzp^2)}{1-\mzp^2/m_W^2} ~.
\ee
Importantly, the sign of the correction $\Delta$ is determined by the relative sign of the $\zprime$ couplings to taus and muons.
The gauged $L_\mu - L_\tau$ unambiguously leads to an enhancement of the $\tau \to \mu \nu_\tau \bar \nu_\mu$ branching ratio. Interestingly, measurements point towards a small positive contribution to the muonic branching ratio of the tau as we now discuss. 

The PDG value for the branching ratio of $\tau\to\mu \nu_\tau \bar \nu_\mu$ reads~\cite{Beringer:1900zz}
\be\label{eq:PDGtau}
\text{BR}(\tau \to \mu \nu_\tau \bar \nu_\mu)_\text{exp} = (17.41 \pm 0.04)\% ~.
\ee
This should be compared to the SM prediction~\cite{Pich:2013lsa}
\be \label{eq:tau_BR_SM}
\text{BR}(\tau \to \mu \nu_\tau \bar \nu_\mu)_\text{SM} = \tau_\tau (5.956 \pm 0.002) \times 10^{11} /s ~.
\ee
The dominant uncertainty on the SM prediction for the branching ratio comes from $\tau_\tau$, the lifetime of the tau.
Combining a very recent result on the tau lifetime from Belle~\cite{Belous:2013dba} with previous measurements at LEP~\cite{Alexander:1996nc,Barate:1997be,Acciarri:2000vq,Abdallah:2003yq} and CLEO~\cite{Balest:1996cr}, results in $\tau_\tau =(290.29 \pm 0.53) \times 10^{-15} s$. Using this value in the SM prediction for BR$(\tau \to \mu \nu_\tau \bar\nu_\mu)$, we find that the experimental value in Eq.~(\ref{eq:PDGtau}) is more than $2\sigma$ {\it above} the SM prediction. Translated into the variable $\Delta$, we obtain
\be
\label{eqn:delta3}
\Delta = (7.0 \pm 3.0) \times 10^{-3}~.
\ee
In Fig.~\ref{fig:constraints1}, the region of parameter space favored by the $\tau$ decay to muons is shown as a green band.

\bigskip
 {$\bullet$ {\bf{ Z coupling to leptons.}}} Loops involving the $\zprime$ also affect the couplings of the SM $\ZZ$ vector-boson to muons, taus and neutrinos. The corresponding branching ratios have been measured very accurately at  LEP and SLC facilities. The corrections to the vector and axial-vector couplings of the $\ZZ$ to leptons are given by
\begin{subequations}
\be
\frac{g_{Ve}}{g_{Ve}^{\rm SM}} &=& \frac{g_{Ae}}{g_{Ae}^{\rm SM}} = 1 ~,\\
\frac{g_{V\mu}}{g_{V\mu}^{\rm SM}} &=& \frac{g_{A\mu}}{g_{A\mu}^{\rm SM}} = \left|1+\frac{(g^\prime)^2}{(4 \pi)^2} K_F(\mzp)\right|~,\\
\frac{g_{V\tau}}{g_{V\tau}^{\rm SM}} &=& \frac{g_{A\tau}}{g_{A\tau}^{\rm SM}} = \left|1+\frac{(g^\prime)^2}{(4 \pi)^2} K_F(\mzp)\right| ~,
\ee
\end{subequations}
where $K_F$ is a loop function that can be found e.g. in~\cite{Haisch:2011up}. 
Out of the three SM neutrinos only the muon-neutrino and tau-neutrino are affected by $Z^\prime$ loops. Therefore, the correction to the $\ZZ$ coupling to neutrinos is effectively given by 
\be
\frac{g_{V\nu}}{g_{V\nu}^{\rm SM}} &=& \frac{g_{A\nu}}{g_{A\nu}^{\rm SM}} = \left|1+ \frac{2}{3} \frac{(g^\prime)^2}{(4 \pi)^2} K_F(\mzp)\right| ~.
\ee
In order to obtain constraints on the mass and coupling of the $Z^\prime$, we combine the experimental results from LEP and SLC~\cite{ALEPH:2005ab} on the $\ZZ$ couplings to all leptons and neutrinos, taking into account the error correlations.
We find the 95\% C.L. constraints depicted in gray in Fig.~\ref{fig:constraints1}. We note also that the constraint on the parameter space would be stronger, if we had a sizable kinetic mixing~\cite{Hook:2011}.

\bigskip
 {$\bullet$ {\boldmath $Z\to 4\ell$} {\bf{searches at the LHC.}}} Both ATLAS and CMS collaborations have reported the measurement of the branching ratio of $\ZZ$ decaying into four charged leptons~\cite{CMS:2012bw,TheATLAScollaboration:2013nha}\footnote{Note that LEP performed the measurement of the cross section of the four-fermion final state arising from the process $e^+ e^-\to \ell^+\ell^- f \bar f$ where $\ell$ is a charged or neutral lepton and $f$ any charged fermion~\cite{aleph4l}. However, as also shown in~\cite{Ma:2001md}, the constraints on the $g^\prime-m_{Z^\prime}$ parameter space coming from this measurement are slightly less stringent than the LHC constraints discussed in the following.}. In particular, the ATLAS analysis~\cite{TheATLAScollaboration:2013nha} has been performed with the full 7+8 TeV LHC data set and it gives BR$(Z\to 4\ell)=(4.2\pm 0.4)10^{-6}$, to be compared to the SM prediction BR$(Z\to 4\ell)=(4.37\pm 0.03)10^{-6}$. Our model gives a positive NP contribution to the process.
 The most important effect comes from the Feynman diagram shown in Fig.~\ref{fig:Z4l}, with an intermediate on-shell $Z'$ boson dominating the rate for $m_{Z'}< m_Z$ (see also~\cite{Harigaya:2013twa} for a recent analysis). 
 
 We have recast the ATLAS analysis in~\cite{TheATLAScollaboration:2013nha}, generating events using MadGraph 5~\cite{Alwall:2011uj}, interfaced with Pythia6.4~\cite{Sjostrand:2006za} for parton showering. Events should have exactly four isolated leptons with the leading three with $p_T > 20,\, 15,\, 8$ GeV, and if the third lepton is an electron it must have $p_T > 10$ GeV. Lepton identification efficiencies have been taken from~\cite{ATLAS:2013nma}. The invariant mass of the opposite sign same flavor (OSSF) lepton pair closest to the $\ZZ$ mass should be $m_1>20$ GeV. The second OSSF lepton invariant mass should be $m_2> 5$ GeV. Finally, the invariant mass of the four lepton system should be close to the $\ZZ$ mass: $80\,{\rm{GeV}}<m_{4\ell}<100\,{\rm{GeV}}$. 
 
NP effects arise only in the four muon bin. In this bin, ATLAS observes 77 events, to be compared to the 78 events expected. To set the bound, we assume a Poisson distribution for the observed events, and we exclude at the 95\% C.L. the benchmarks that predict more than 94 events in the four muon bin. The region on the left of the dashed black line in Fig.~\ref{fig:constraints1} is excluded by the ATLAS analysis. As we can note from the figure, the region favored by $(g-2)_\mu$ has been almost fully probed by LHC measurements of $\ZZ$ to four leptons.

\begin{figure}[t]
\centering
\includegraphics[width=0.4\textwidth]{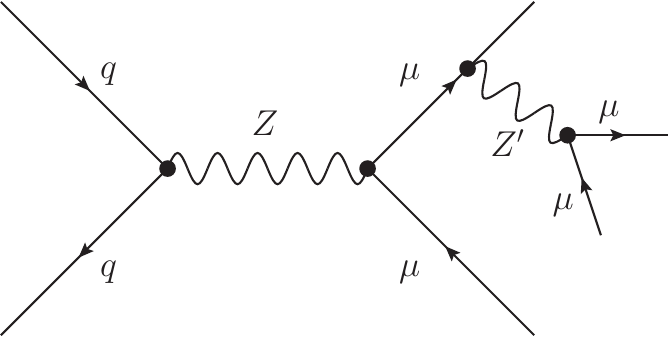}%
\caption{The main NP contribution to the $Z\to 4\ell$ process at the LHC.}
\label{fig:Z4l}
\end{figure}

\bigskip


\begin{figure}[t]
\centering
\includegraphics[width=0.4\textwidth]{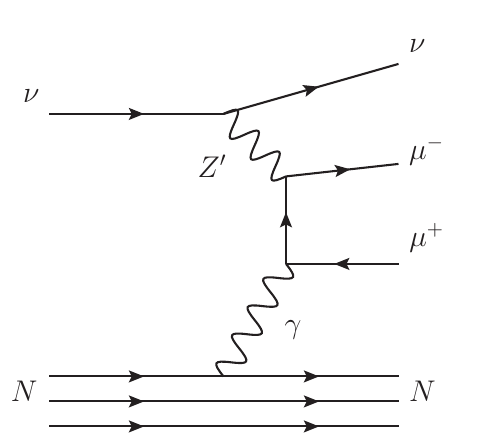}%
\caption{The leading order contribution of the $\zprime$ to neutrino trident production. This diagram interferes constructively (destructively) with the corresponding SM diagram involving a $W$-boson ($\ZZ$-boson).}
\label{fig:trident_diagram}
\end{figure}

{$\bullet$ {\bf Neutrino trident production.}} In the last part of this section, we present a powerful new constraint on the $L_\mu-L_\tau$ current coming from measurements of neutrino trident production, {\it{i.e.}} the production of a muon anti-muon pair in the scattering of muon neutrinos in the Coulomb field of a target nucleus. The leading contribution of the $\zprime$ to such a process is shown in Fig.~\ref{fig:trident_diagram}. This diagram interferes with the SM contribution involving similar diagrams, but with the $W$ and $\ZZ$ bosons instead of the $\zprime$. In the SM, the contribution from the $\ZZ$-boson is smaller than the one of the $W$-boson and comes with an opposite sign 
that leads to destructive interference~\cite{Belusevic:1987cw}. The $\zprime$ coupling to both muons and muon-neutrinos has the same sign and the $Z^\prime$ contribution interferes constructively (destructively) with the $W$-boson ($\ZZ$-boson), leading therefore to an enhancement of the trident production. Working in the approximation of a heavy $\zprime$, where the leptonic 4-fermion operator is $ (\gp)^2\left( \bar \mu \gamma_\alpha \mu \right) \left( \bar \nu \gamma^\alpha P_L \nu \right)/\mzp^2$\footnote{We estimate that the description of the $Z^\prime$ contribution by an effective 4-fermion operator is accurate as long as $\mzp \gtrsim 10$~GeV. A detailed analysis of neutrino trident production in the presence of a lighter $Z^\prime$ will be presented elsewhere~\cite{Altmannshofer:2014pba}.}, the ratio of the total trident cross-section to the SM prediction is given by
\be \label{eq:trident_correction}
\frac{\sigma}{\sigma_\text{SM}} \simeq \frac{1 + \left( 1 + 4 s_W^2 + 2v^2/v_\phi^2 \right)^2 }{1 + \left( 1 + 4 s_W^2 \right)^2} ~.
\ee 

Neutrino trident production has been observed by three experiments: the first positive results came from the CHARM-II collaboration~\cite{Geiregat:1990gz}; the next measurement was by the CCFR collaboration~\cite{Mishra:1991bv}, 
further confirmed by the NuTeV collaboration~\cite{Adams:1998yf}.
Combining the measured cross sections with the corresponding SM predictions we find
\begin{align}
\sigma_{\rm{CHARM-II}}/\sigma_{\rm{SM}} = 1.58 \pm 0.57  \label{eq:CHARM}
 ~,\\
 \sigma_{\rm{CCFR}}/\sigma_{\rm{SM}} = 0.82 \pm 0.28  \label{eq:CCFR}
 ~, \\
 \sigma_{\rm{NuTeV}}/\sigma_{\rm{SM}} = 0.67 \pm 0.27  \label{eq:NuTeV}
 ~.
\end{align}
A weighted average gives
\be
\label{eq:trident_WA}
 \sigma_\text{exp}/\sigma_\text{SM} = 0.83 \pm 0.18 ~,
\ee
which leaves only little room for positive NP contributions. Combining Eq.~(\ref{eq:trident_WA}) with~(\ref{eq:trident_correction}) we find
\be \label{eq:trident_bound}
v_\phi \gtrsim 750 ~\text{GeV}~.
\ee
A later published paper of NuTeV~\cite{Adams:1999mn} reports a much less stringent bound than in~\cite{Adams:1998yf}. Combining this bound with the ones from the CHARM-II and CCFR collaborations, we get $\sigma_\text{exp}/\sigma_\text{SM}=0.95 \pm 0.25$, that is only slightly weaker than the bound we show in Eq. (\ref{eq:trident_WA}), not changing therefore the nature of our conclusions. 
 
The bound in (\ref{eq:trident_bound}) completely excludes an explanation of the $(g-2)_\mu$ anomaly for the $\mzp\gtrsim 10$ GeV region we consider in this paper. 
The constraint coming from Eq.~(\ref{eq:trident_WA}) as well as the individual constraints from Eqs.~(\ref{eq:CHARM}) and~(\ref{eq:CCFR}) are shown by the red lines in Fig.~\ref{fig:constraints1} in the $\mzp$ - $g^\prime$ plane.

\bigskip
{$\bullet$ {\bf Final remarks.}} Fig.~\ref{fig:constraints1} is a summary of all the leptonic constraints on $L_\mu - L_\tau$ discussed in this section. Remarkably, a major part of the parameter space relevant for the $B\rightarrow K^* \mu^+\mu^-$ anomaly, and {\emph {all}} of the parameter space relevant for the muon $g-2$ anomaly, is probed by the observation of neutrino trident production. The enormous potential of this process in providing full coverage of the parameter space strongly motivates future experiments looking to measure this process more precisely. 

Finally, using the lower bound on the VEV from the neutrino tridents, we can predict a minimum effect in $B_s$ mixing, if the $Z^\prime$ is to explain the $B \to K^* \mu^+\mu^-$ anomaly. We find that the mass difference in the $B_s$ system, $\Delta M_s$ is affected by at least 3\%, and the effect grows quadratically with $v_\Phi$. While a 3\% effect in $\Delta M_s$ is well within the uncertainty of the SM prediction, 
for generic  values of the Yukawa couplings one should expect an effect of the same order also in the theoretically clean $B_s$ mixing phase, which should be detectable with an LHCb upgrade~\cite{Bediaga:2012py}.
The expected effects in $B_s$ mixing are indicated in the white region of Fig.~\ref{fig:constraints1} by the dotted contours.

\section{Outlook and Conclusions}
\label{sec:outlook}

This work was devoted to a comprehensive study of a model with a $Z'$ vector-boson that couples to leptons through the $L_\mu-L_\tau$ portal, and to quarks through general effective couplings. Our goal was to determine whether such a model 
yields a plausible explanation for the recent discrepancy shown by the LHCb collaboration in angular distributions of 
the $B\to K^*\mu^+\mu^-$ decay products. We conclude that such an explanation is viable, and it is such that future measurements in the 
high-energy and high-intensity frontiers may reveal further deviations from the SM tied to the manifestations 
of this new vector-boson.
Unlike models based on a $Z'$ that couples with full strength to all leptons and quarks, the model we consider in this paper 
is well-hidden. In contradistinction to most of the $Z'$ proposals made in connection with the LHCb discrepancy, which envision a $Z'$ above $\gtrsim 3 $ TeV,  the mass of the vector-boson considered in this work can be very low, possibly well below the electroweak scale! 
While a variety of UV-completions are possible for the coupling of $Z'$ to quarks, we have chosen one with vector-like quarks in the multi-TeV mass scale. While this model can hardly be imagined to be the final word, it does offer a general and consistent framework within which it is possible to discuss the different low-energy constraints and structures likely to emerge in more refined constructions. 

Among the leptonic observables, we have identified two particular processes which result in powerful constraints 
on the parameter space of the model: the $\ZZ$ decay to four muons and the neutrino trident production. In particular, we find that the tentative explanation of the $(g-2)_\mu$ discrepancy in this model is fully ruled out by the latter process, at least for multi-GeV and heavier $Z^\prime$. While in this work we have applied it to the $L_\mu-L_\tau$ portal, it is absolutely clear that neutrino trident production is immediately relevant to other models that appeal to $Z'$ coupled to leptons via any current that contains $L_\mu$ (such as {\em e.g.} total lepton number). Generalizing this constraint to other models and extending it to a wider range of the $Z'$ mass is the subject of our upcoming work \cite{Altmannshofer:2014pba}.

\begin{acknowledgments}
We would like to thank P. Langacker and A. Ritz for useful discussions. We acknowledge the useful correspondence with K. Harigaya, T. Igari, M. M. Nojiri, M. Takeuchi and K. Tobe. 
We would like to thank C. Bobeth, A. Buras, A. Celis and M. Jung for pointing out a typo in Eq.~(\ref{eq:19c}) in earlier versions of this paper.
WA and SG acknowledge the kind hospitality of the Particle Theory Group at the University of Victoria where part of this work was done. IY would like to extend a similar acknowledgment to the kind hospitality of the Johns Hopkins University. The research of WA was supported by the John Templeton Foundation. Research at Perimeter Institute is supported by the Government of Canada through Industry Canada and by the Province of Ontario through the Ministry of Economic Development \& Innovation. 
\end{acknowledgments}

\bibliography{mu-tau_bib}

\providecommand{\href}[2]{#2}\begingroup\raggedright\begin{thebibliography}{10}

\bibitem{Aaij:2013qta}
{\bf LHCb} Collaboration, R.~Aaij {\em et al.}, ``{Measurement of form-factor
  independent observables in the decay $B^{0} \to K^{*0} \mu^+ \mu^-$},''
  \href{http://dx.doi.org/10.1103/PhysRevLett.111.191801}{{\em Phys.Rev.Lett.}
  {\bf 111} (2013)  191801},
\href{http://arxiv.org/abs/1308.1707}{{\tt arXiv:1308.1707 [hep-ex]}}.

\bibitem{Descotes-Genon:2013wba}
S.~Descotes-Genon, J.~Matias, and J.~Virto, ``{Understanding the $B \to
  K^*\mu^+\mu^-$ Anomaly},''
  \href{http://dx.doi.org/10.1103/PhysRevD.88.074002}{{\em Phys.Rev.} {\bf D88}
  (2013)  074002},
\href{http://arxiv.org/abs/1307.5683}{{\tt arXiv:1307.5683 [hep-ph]}}.

\bibitem{Altmannshofer:2013foa}
W.~Altmannshofer and D.~M. Straub, ``{New physics in $B \to K^*\mu\mu$?},''
  \href{http://dx.doi.org/10.1140/epjc/s10052-013-2646-9}{{\em Eur.Phys.J.}
  {\bf C73} (2013)  2646},
\href{http://arxiv.org/abs/1308.1501}{{\tt arXiv:1308.1501 [hep-ph]}}.

\bibitem{Beaujean:2013soa}
F.~Beaujean, C.~Bobeth, and D.~van Dyk, ``{Comprehensive Bayesian Analysis of
  Rare (Semi)leptonic and Radiative $B$ Decays},''
\href{http://arxiv.org/abs/1310.2478}{{\tt arXiv:1310.2478 [hep-ph]}}.

\bibitem{Hurth:2013ssa}
T.~Hurth and F.~Mahmoudi, ``{On the LHCb anomaly in $B \to K^* l^+ l^-$},''
\href{http://arxiv.org/abs/1312.5267}{{\tt arXiv:1312.5267 [hep-ph]}}.

\bibitem{Gauld:2013qba}
R.~Gauld, F.~Goertz, and U.~Haisch, ``{On minimal $Z^\prime$ explanations of
  the $B\to K^*\mu^+\mu^-$ anomaly},''
  \href{http://dx.doi.org/10.1103/PhysRevD.89.015005}{{\em Phys.Rev.} {\bf D89}
  (2014)  015005},
\href{http://arxiv.org/abs/1308.1959}{{\tt arXiv:1308.1959 [hep-ph]}}.

\bibitem{Buras:2013qja}
A.~J. Buras and J.~Girrbach, ``{Left-handed $Z^\prime$ and $Z$ FCNC quark
  couplings facing new $b \to s \mu^+ \mu^-$ data},''
  \href{http://dx.doi.org/10.1007/JHEP12(2013)009}{{\em JHEP} {\bf 1312} (2013)
   009},
\href{http://arxiv.org/abs/1309.2466}{{\tt arXiv:1309.2466 [hep-ph]}}.

\bibitem{Gauld:2013qja}
R.~Gauld, F.~Goertz, and U.~Haisch, ``{An explicit $Z^\prime$-boson explanation
  of the $B \to K^* \mu^+ \mu^-$ anomaly},''
  \href{http://dx.doi.org/10.1007/JHEP01(2014)069}{{\em JHEP} {\bf 1401} (2014)
   069},
\href{http://arxiv.org/abs/1310.1082}{{\tt arXiv:1310.1082 [hep-ph]}}.

\bibitem{Buras:2013dea}
A.~J. Buras, F.~De~Fazio, and J.~Girrbach, ``{331 models facing new $b \to s
  \mu^+ \mu^-$ data},''
\href{http://arxiv.org/abs/1311.6729}{{\tt arXiv:1311.6729 [hep-ph]}}.

\bibitem{Datta:2013kja}
A.~Datta, M.~Duraisamy, and D.~Ghosh, ``{Explaining the $B \to K^\ast \mu^+
  \mu^-$ anomaly with scalar interactions},''
\href{http://arxiv.org/abs/1310.1937}{{\tt arXiv:1310.1937 [hep-ph]}}.

\bibitem{Mahmoudi:2014mja}
F.~Mahmoudi, S.~Neshatpour, and J.~Virto, ``{$B \to K^{*} \mu^{+} \mu^{-}$
  optimised observables in the MSSM},''
\href{http://arxiv.org/abs/1401.2145}{{\tt arXiv:1401.2145 [hep-ph]}}.

\bibitem{He:1990pn}
X.~He, G.~C. Joshi, H.~Lew, and R.~Volkas, ``{New $Z^\prime$ penomenology},''
\href{http://dx.doi.org/10.1103/PhysRevD.43.22}{{\em Phys.Rev.} {\bf D43}
  (1991)  22--24}.

\bibitem{He:1991qd}
X.-G. He, G.~C. Joshi, H.~Lew, and R.~Volkas, ``{Simplest $Z^\prime$ model},''
\href{http://dx.doi.org/10.1103/PhysRevD.44.2118}{{\em Phys.Rev.} {\bf D44}
  (1991)  2118--2132}.

\bibitem{Baek:2001kca}
S.~Baek, N.~Deshpande, X.~He, and P.~Ko, ``{Muon anomalous $g-2$ and gauged
  $L_\mu - L_\tau$ models},''
  \href{http://dx.doi.org/10.1103/PhysRevD.64.055006}{{\em Phys.Rev.} {\bf D64}
  (2001)  055006},
\href{http://arxiv.org/abs/hep-ph/0104141}{{\tt arXiv:hep-ph/0104141
  [hep-ph]}}.

\bibitem{Ma:2001md}
E.~Ma, D.~Roy, and S.~Roy, ``{Gauged $L_\mu - L_\tau$ with large muon anomalous
  magnetic moment and the bimaximal mixing of neutrinos},''
  \href{http://dx.doi.org/10.1016/S0370-2693(01)01428-9}{{\em Phys.Lett.} {\bf
  B525} (2002)  101--106},
\href{http://arxiv.org/abs/hep-ph/0110146}{{\tt arXiv:hep-ph/0110146
  [hep-ph]}}.

\bibitem{Salvioni:2009jp}
E.~Salvioni, A.~Strumia, G.~Villadoro, and F.~Zwirner, ``{Non-universal minimal
  $Z^\prime$ models: present bounds and early LHC reach},''
  \href{http://dx.doi.org/10.1007/JHEP03(2010)010}{{\em JHEP} {\bf 1003} (2010)
   010},
\href{http://arxiv.org/abs/0911.1450}{{\tt arXiv:0911.1450 [hep-ph]}}.

\bibitem{Heeck:2011wj}
J.~Heeck and W.~Rodejohann, ``{Gauged $L_\mu$ - $L_\tau$ Symmetry at the
  Electroweak Scale},''
  \href{http://dx.doi.org/10.1103/PhysRevD.84.075007}{{\em Phys.Rev.} {\bf D84}
  (2011)  075007},
\href{http://arxiv.org/abs/1107.5238}{{\tt arXiv:1107.5238 [hep-ph]}}.

\bibitem{Feng:2012jn}
W.-Z. Feng, P.~Nath, and G.~Peim, ``{Cosmic Coincidence and Asymmetric Dark
  Matter in a Stueckelberg Extension},''
  \href{http://dx.doi.org/10.1103/PhysRevD.85.115016}{{\em Phys.Rev.} {\bf D85}
  (2012)  115016},
\href{http://arxiv.org/abs/1204.5752}{{\tt arXiv:1204.5752 [hep-ph]}}.

\bibitem{Harigaya:2013twa}
K.~Harigaya, T.~Igari, M.~M. Nojiri, M.~Takeuchi, and K.~Tobe, ``{Muon $g-2$
  and LHC phenomenology in the $L_\mu-L_\tau$ gauge symmetric model},''
\href{http://arxiv.org/abs/1311.0870}{{\tt arXiv:1311.0870 [hep-ph]}}.

\bibitem{Fox:2011qd}
P.~J. Fox, J.~Liu, D.~Tucker-Smith, and N.~Weiner, ``{An Effective
  $Z^\prime$},'' \href{http://dx.doi.org/10.1103/PhysRevD.84.115006}{{\em
  Phys.Rev.} {\bf D84} (2011)  115006},
\href{http://arxiv.org/abs/1104.4127}{{\tt arXiv:1104.4127 [hep-ph]}}.

\bibitem{Langacker:2008yv}
P.~Langacker, ``{The Physics of Heavy $Z^\prime$ Gauge Bosons},''
  \href{http://dx.doi.org/10.1103/RevModPhys.81.1199}{{\em Rev.Mod.Phys.} {\bf
  81} (2009)  1199--1228},
\href{http://arxiv.org/abs/0801.1345}{{\tt arXiv:0801.1345 [hep-ph]}}.

\bibitem{Altmannshofer:2014pba}
W.~Altmannshofer, S.~Gori, M.~Pospelov, and I.~Yavin, ``{Neutrino Trident
  Production: A Powerful Probe of New Physics with Neutrino Beams},''
  \href{http://dx.doi.org/10.1103/PhysRevLett.113.091801}{{\em Phys. Rev.
  Lett.} {\bf 113} (2014)  091801},
\href{http://arxiv.org/abs/1406.2332}{{\tt arXiv:1406.2332 [hep-ph]}}.

\bibitem{Carone:2013uh}
C.~D. Carone, ``{Flavor-Nonuniversal Dark Gauge Bosons and the Muon g-2},''
  \href{http://dx.doi.org/10.1016/j.physletb.2013.03.011}{{\em Phys.Lett.} {\bf
  B721} (2013)  118--122},
\href{http://arxiv.org/abs/1301.2027}{{\tt arXiv:1301.2027 [hep-ph]}}.

\bibitem{Buras:2012jb}
A.~J. Buras, F.~De~Fazio, and J.~Girrbach, ``{The Anatomy of Z' and Z with
  Flavour Changing Neutral Currents in the Flavour Precision Era},''
  \href{http://dx.doi.org/10.1007/JHEP02(2013)116}{{\em JHEP} {\bf 1302} (2013)
   116},
\href{http://arxiv.org/abs/1211.1896}{{\tt arXiv:1211.1896 [hep-ph]}}.

\bibitem{Isidori:2010kg}
G.~Isidori, Y.~Nir, and G.~Perez, ``{Flavor Physics Constraints for Physics
  Beyond the Standard Model},''
  \href{http://dx.doi.org/10.1146/annurev.nucl.012809.104534}{{\em
  Ann.Rev.Nucl.Part.Sci.} {\bf 60} (2010)  355},
\href{http://arxiv.org/abs/1002.0900}{{\tt arXiv:1002.0900 [hep-ph]}}.

\bibitem{Butler:2013kdw}
{\bf Quark Flavor Physics Working Group} Collaboration, J.~Butler {\em et al.},
  ``{Report of the Quark Flavor Physics Working Group},''
\href{http://arxiv.org/abs/1311.1076}{{\tt arXiv:1311.1076 [hep-ex]}}.

\bibitem{Aushev:2010bq}
T.~Aushev, W.~Bartel, A.~Bondar, J.~Brodzicka, T.~Browder, {\em et al.},
  ``{Physics at Super B Factory},''
\href{http://arxiv.org/abs/1002.5012}{{\tt arXiv:1002.5012 [hep-ex]}}.

\bibitem{Kozhuharov:2013oca}
{\bf NA62} Collaboration, V.~Kozhuharov, ``{Rare K Decays: Present and
  perspectives with NA62},''
\href{http://arxiv.org/abs/1305.2840}{{\tt arXiv:1305.2840 [hep-ex]}}.

\bibitem{Aad:2012ij}
{\bf ATLAS} Collaboration, G.~Aad {\em et al.}, ``{A search for flavour
  changing neutral currents in top-quark decays in $pp$ collision data
  collected with the ATLAS detector at $\sqrt{s}=7$ TeV},''
  \href{http://dx.doi.org/10.1007/JHEP09(2012)139}{{\em JHEP} {\bf 1209} (2012)
   139},
\href{http://arxiv.org/abs/1206.0257}{{\tt arXiv:1206.0257 [hep-ex]}}.

\bibitem{Chatrchyan:2013nwa}
{\bf CMS} Collaboration, S.~Chatrchyan {\em et al.}, ``{Search for
  flavor-changing neutral currents in top-quark decays $t \to Zq$ in pp
  collisions at $\sqrt{s}$=8 TeV},''
\href{http://arxiv.org/abs/1312.4194}{{\tt arXiv:1312.4194 [hep-ex]}}.

\bibitem{ATLAS:2013hta}
{\bf ATLAS} Collaboration, ``{Physics at a High-Luminosity LHC with ATLAS},''
\href{http://arxiv.org/abs/1307.7292}{{\tt arXiv:1307.7292 [hep-ex]}}.

\bibitem{CMS:2013xfa}
{\bf CMS} Collaboration, ``{Projected Performance of an Upgraded CMS Detector
  at the LHC and HL-LHC: Contribution to the Snowmass Process},''
\href{http://arxiv.org/abs/1307.7135}{{\tt arXiv:1307.7135 [hep-ex]}}.

\bibitem{Pospelov:2008zw}
M.~Pospelov, ``{Secluded U(1) below the weak scale},''
  \href{http://dx.doi.org/10.1103/PhysRevD.80.095002}{{\em Phys.Rev.} {\bf D80}
  (2009)  095002},
\href{http://arxiv.org/abs/0811.1030}{{\tt arXiv:0811.1030 [hep-ph]}}.

\bibitem{Jegerlehner:2009ry}
F.~Jegerlehner and A.~Nyffeler, ``{The Muon $g-2$},''
  \href{http://dx.doi.org/10.1016/j.physrep.2009.04.003}{{\em Phys.Rept.} {\bf
  477} (2009)  1--110},
\href{http://arxiv.org/abs/0902.3360}{{\tt arXiv:0902.3360 [hep-ph]}}.

\bibitem{Beringer:1900zz}
{\bf Particle Data Group} Collaboration, J.~Beringer {\em et al.}, ``{Review of
  Particle Physics (RPP)},''
  \href{http://dx.doi.org/10.1103/PhysRevD.86.010001}{{\em Phys.Rev.} {\bf D86}
  (2012)  010001}.
{and 2013 partial update for the 2014 edition}.

\bibitem{Pich:2013lsa}
A.~Pich, ``{Precision Tau Physics},''
\href{http://arxiv.org/abs/1310.7922}{{\tt arXiv:1310.7922 [hep-ph]}}.

\bibitem{Belous:2013dba}
{\bf Belle} Collaboration, K.~Belous {\em et al.}, ``{Measurement of the
  $\tau$-lepton lifetime at Belle},''
\href{http://arxiv.org/abs/1310.8503}{{\tt arXiv:1310.8503 [hep-ex]}}.

\bibitem{Alexander:1996nc}
{\bf OPAL} Collaboration, G.~Alexander {\em et al.}, ``{Improved measurement of
  the lifetime of the tau lepton},''
\href{http://dx.doi.org/10.1016/0370-2693(96)00255-9}{{\em Phys.Lett.} {\bf
  B374} (1996)  341--350}.

\bibitem{Barate:1997be}
{\bf ALEPH} Collaboration, R.~Barate {\em et al.}, ``{Updated measurement of
  the tau lepton lifetime},''
  \href{http://dx.doi.org/10.1016/S0370-2693(97)01116-7}{{\em Phys.Lett.} {\bf
  B414} (1997)  362--372},
\href{http://arxiv.org/abs/hep-ex/9710026}{{\tt arXiv:hep-ex/9710026
  [hep-ex]}}.

\bibitem{Acciarri:2000vq}
{\bf L3} Collaboration, M.~Acciarri {\em et al.}, ``{Measurement of the
  lifetime of the $\tau$ lepton},''
  \href{http://dx.doi.org/10.1016/S0370-2693(00)00347-6}{{\em Phys.Lett.} {\bf
  B479} (2000)  67--78},
\href{http://arxiv.org/abs/hep-ex/0003023}{{\tt arXiv:hep-ex/0003023
  [hep-ex]}}.

\bibitem{Abdallah:2003yq}
{\bf DELPHI} Collaboration, J.~Abdallah {\em et al.}, ``{A Precise measurement
  of the tau lifetime},''
  \href{http://dx.doi.org/10.1140/epjc/s2004-01953-7}{{\em Eur.Phys.J.} {\bf
  C36} (2004)  283--296},
\href{http://arxiv.org/abs/hep-ex/0410010}{{\tt arXiv:hep-ex/0410010
  [hep-ex]}}.

\bibitem{Balest:1996cr}
{\bf CLEO} Collaboration, R.~Balest {\em et al.}, ``{Measurement of the tau
  lepton lifetime},''
\href{http://dx.doi.org/10.1016/S0370-2693(96)01163-X}{{\em Phys.Lett.} {\bf
  B388} (1996)  402--408}.

\bibitem{Haisch:2011up}
U.~Haisch and S.~Westhoff, ``{Massive Color-Octet Bosons: Bounds on Effects in
  Top-Quark Pair Production},''
  \href{http://dx.doi.org/10.1007/JHEP08(2011)088}{{\em JHEP} {\bf 1108} (2011)
   088},
\href{http://arxiv.org/abs/1106.0529}{{\tt arXiv:1106.0529 [hep-ph]}}.

\bibitem{ALEPH:2005ab}
{\bf ALEPH, DELPHI, L3, OPAL, SLD, LEP Electroweak Working Group, SLD
  Electroweak Group, SLD Heavy Flavour Group} Collaboration, S.~Schael {\em et
  al.}, ``{Precision electroweak measurements on the $Z$ resonance},''
  \href{http://dx.doi.org/10.1016/j.physrep.2005.12.006}{{\em Phys.Rept.} {\bf
  427} (2006)  257--454},
\href{http://arxiv.org/abs/hep-ex/0509008}{{\tt arXiv:hep-ex/0509008
  [hep-ex]}}.

\bibitem{Hook:2011}
A.~Hook, E.~Izaguirre, and J.~G. Wacker, ``{Model Independent Bounds on Kinetic
  Mixing},'' \href{http://dx.doi.org/10.1155/2011/859762}{{\em Adv.High Energy
  Phys.} {\bf 2011} (2011)  859762},
\href{http://arxiv.org/abs/1006.0973}{{\tt arXiv:1006.0973 [hep-ph]}}.

\bibitem{CMS:2012bw}
{\bf CMS} Collaboration, S.~Chatrchyan {\em et al.}, ``{Observation of Z decays
  to four leptons with the CMS detector at the LHC},''
  \href{http://dx.doi.org/10.1007/JHEP12(2012)034}{{\em JHEP} {\bf 1212} (2012)
   034},
\href{http://arxiv.org/abs/1210.3844}{{\tt arXiv:1210.3844 [hep-ex]}}.

\bibitem{TheATLAScollaboration:2013nha}
{\bf ATLAS} Collaboration, ``{ATLAS measurements of the 7 and 8 TeV cross
  sections for $Z\rightarrow 4\ell$ in pp collisions},''.
ATLAS-CONF-2013-055.

\bibitem{aleph4l}
{\bf ALEPH} Collaboration, D.~Buskulic {\em et al.}, ``Study of the
  four-fermion final state at the z resonance,''
  \href{http://dx.doi.org/10.1007/BF01496576}{{\em Zeitschrift f{\"u}r Physik C
  Particles and Fields} {\bf 66} (1995)  }.

\bibitem{Alwall:2011uj}
J.~Alwall, M.~Herquet, F.~Maltoni, O.~Mattelaer, and T.~Stelzer, ``{MadGraph 5
  : Going Beyond},'' \href{http://dx.doi.org/10.1007/JHEP06(2011)128}{{\em
  JHEP} {\bf 1106} (2011)  128},
\href{http://arxiv.org/abs/1106.0522}{{\tt arXiv:1106.0522 [hep-ph]}}.

\bibitem{Sjostrand:2006za}
T.~Sjostrand, S.~Mrenna, and P.~Z. Skands, ``{PYTHIA 6.4 Physics and Manual},''
  \href{http://dx.doi.org/10.1088/1126-6708/2006/05/026}{{\em JHEP} {\bf 0605}
  (2006)  026},
\href{http://arxiv.org/abs/hep-ph/0603175}{{\tt arXiv:hep-ph/0603175
  [hep-ph]}}.

\bibitem{ATLAS:2013nma}
{\bf ATLAS} Collaboration, ``{Measurements of the properties of the Higgs-like
  boson in the four lepton decay channel with the ATLAS detector using 25
  fb$^{−1}$ of proton-proton collision data},''.
ATLAS-CONF-2013-013.

\bibitem{Belusevic:1987cw}
R.~Belusevic and J.~Smith, ``{$W$ - $Z$ Interference in Neutrino - Nucleus
  Scattering},''
\href{http://dx.doi.org/10.1103/PhysRevD.37.2419}{{\em Phys.Rev.} {\bf D37}
  (1988)  2419}.

\bibitem{Geiregat:1990gz}
{\bf CHARM-II} Collaboration, D.~Geiregat {\em et al.}, ``{First observation of
  neutrino trident production},''
\href{http://dx.doi.org/10.1016/0370-2693(90)90146-W}{{\em Phys.Lett.} {\bf
  B245} (1990)  271--275}.

\bibitem{Mishra:1991bv}
{\bf CCFR} Collaboration, S.~Mishra {\em et al.}, ``{Neutrino tridents and
  $W$-$Z$ interference},''
\href{http://dx.doi.org/10.1103/PhysRevLett.66.3117}{{\em Phys.Rev.Lett.} {\bf
  66} (1991)  3117--3120}.

\bibitem{Adams:1998yf}
{\bf NuTeV} Collaboration, T.~Adams {\em et al.}, ``{Neutrino trident
  production from NuTeV},''
\href{http://arxiv.org/abs/hep-ex/9811012}{{\tt arXiv:hep-ex/9811012
  [hep-ex]}}.

\bibitem{Adams:1999mn}
{\bf NuTeV} Collaboration, T.~Adams {\em et al.}, ``{Evidence for diffractive
  charm production in muon-neutrino Fe and anti-muon-neutrino Fe scattering at
  the Tevatron},'' \href{http://dx.doi.org/10.1103/PhysRevD.61.092001}{{\em
  Phys. Rev.} {\bf D61} (2000)  092001},
\href{http://arxiv.org/abs/hep-ex/9909041}{{\tt arXiv:hep-ex/9909041
  [hep-ex]}}.

\bibitem{Bediaga:2012py}
{\bf LHCb} Collaboration, R.~Aaij {\em et al.}, ``{Implications of LHCb
  measurements and future prospects},''
  \href{http://dx.doi.org/10.1140/epjc/s10052-013-2373-2}{{\em Eur.Phys.J.}
  {\bf C73} (2013)  2373},
\href{http://arxiv.org/abs/1208.3355}{{\tt arXiv:1208.3355 [hep-ex]}}.

\end{thebibliography}\endgroup

\end{document}